\documentclass[aps,prl,twocolumn,groupedaddress,showpacs]{revtex4}

\usepackage{graphicx}
\usepackage[usenames]{color}%

\begin{document}


\title{New class of small amplitude low-field magnetoresistance oscillation in unidirectional lateral superlattice: Geometric resonance of Bragg-reflected cyclotron orbit}


\author{Akira Endo}
\email[]{akrendo@issp.u-tokyo.ac.jp}
\author{Yasuhiro Iye}
\affiliation{Institute for Solid State Physics, University of Tokyo}


\date{\today}

\begin{abstract}
We have uncovered a new class of small amplitude magnetoresistance oscillation in unidirectional lateral superlattice (ULSL). The oscillation is observed in a low-field regime, typically $|B|$$\alt$ 0.03 T, as small undulation on top of well-known positive magnetoresistance background. Positions of maxima of the oscillation shift to lower field side with the increase of the electron concentration $n_e$ roughly proportionally to $n_e$$^{-1/2}$, and also with the increase of period $a$ of ULSL samples. The oscillation is attributed to commensurability between the period $a$ and the width of open orbits originating from the miniband structure.
\end{abstract}


\maketitle

Unidirectional lateral superlattice (ULSL) --- two-dimensional electron gas (2DEG) subjected to one-dimensional periodic modulation --- has been perceived, since the very early stage of its experimental realization some fifteen years ago \cite{Weiss89}, to exhibit two characteristic magnetotransport features: positive magnetoresistance (PMR) emanating from zero magnetic field \cite{Weiss89,Beton90P}, and commensurability oscillation (CO) originating from geometric resonance between the cyclotron radius $R_c$ and the period $a$ of ULSL \cite{Weiss89,Winkler89}. These features can be basically understood as an outcome of the properties of semiclassical electron orbit under both perpendicular magnetic field $B$ and periodic potential landscape. Although the ingenious concept of superlattice was originally invented aiming at the possibility of designing artificial band structure (miniband) that possesses length and energy scale quite different from that of natural crystals \cite{Esaki70}, both PMR and CO do not necessarily require miniband structure for their explanation. It was not until quite recently that clear evidence of miniband structure is observed in ULSL \cite{Deutschmann01}. In the present paper, we report a new type of magnetoresistance oscillation having very small amplitude, in the low-field regime dominated by PMR\@. The oscillation is attributed to geometric resonance of open orbits resulting from the miniband structure. The interesting phenomenon, which can in principle take place in metals or any other materials that contain open orbits, has been brought within the access of detailed experimental investigation in ULSL by virtue of large man-made lattice constants $a$ and the controllability of the Fermi wavenumber $k_\mathrm{F}$.

\begin{table}
\caption{Properties of Samples\protect\footnote{at 1.4 K} \label{Samples}}
\begin{ruledtabular}
\begin{tabular}{ccccccc}
  & $a$ (nm) & $V_0$ (meV) & $\mu$ (m$^2$/Vs) & $n_e$ (10$^{15}$ m$^{-2})$ & $\eta$=$V_0/E_F$ \\
\hline
\#1 & 184 & $\sim$0.30 & 75$-$108 & 2.0$-$3.1 & 0.028$-$0.044 \\
\#2 & 184 & $\sim$0.27 & 75$-$119 & 2.0$-$3.0 & 0.025$-$0.038 \\
\#3 & 161 & $\sim$0.20 & 73$-$110 & 2.0$-$3.1 & 0.018$-$0.028 \\
\#4 & 138 & $\sim$0.10 & 73$-$91 & 2.0$-$2.9 & 0.010$-$0.014 \\
\end{tabular}
\end{ruledtabular}
\end{table}

\begin{figure}[bt]
\includegraphics[bbllx=20,bblly=280,bburx=800,bbury=800,width=8.5cm]{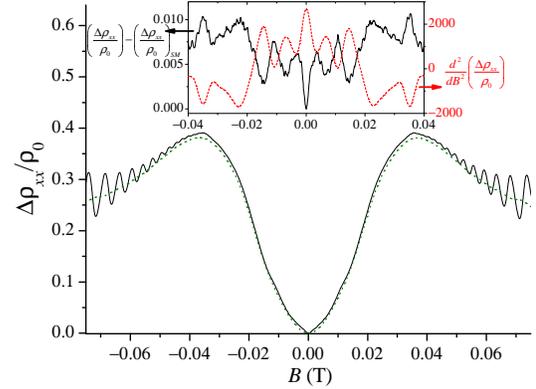}%
\caption{\label{SmoothBG}Magnetoresistance of sample \#2 at 1.4 K, showing PMR ($|B|$$\alt$0.04 T) and CO ($|B|$$\agt$0.04 T). New oscillation is also discernible as small undulation overlaid on PMR at $|B|$$\alt$0.03 T. Inset: New oscillation highlighted either by subtracting slowly varying background (shown by dotted curve in the main panel), or by taking second derivative with respect to $B$.}
\end{figure}

Four ULSL samples with different periods were prepared from the same GaAs/Al$_{0.3}$Ga$_{0.7}$As single heterostructure wafer as tabulated in Table \ref{Samples} (Hall bars with length and width 64$\times$37 $\mu$m$^2$ for \#1, 2 and 44$\times$16 $\mu$m$^2$ for \#3, 4). Potential modulation was introduced by strain-induced piezoelectric effect \cite{Skuras97} employing the surface grating made of electron-beam resist \cite{Endo00e}. Grating was placed perpendicular to the direction of current ($x$ direction) so that the current flew across the modulation. The modulation amplitudes $V_0$ were measured by fitting CO to a formula that takes into account the decay of amplitude in the lower field due to scattering (see \cite{Endo00e} for detail). Since the depth of 2DEG plane ($d$=90 nm) is comparable to the period $a$, $V_0$ is strongly dependent on $a$. The $V_0$ of sample \#2 was intentionally made smaller than that of sample \#1 by using patterned grating that has, along the grating line, 46 nm wide rifts in every $a^\prime$=575 nm designed to partially relax the strain \cite{Endo04P}. Measurements were performed at either 1.4 K or 4.2 K\@. To see the dependence of the oscillation on $k_\mathrm{F}$=$\sqrt{2\pi n_e}$, electron density $n_e$ was progressively increased by step-by-step LED illumination, with concomitant increase of the mobility $\mu$. (Samples \#3 and \#4 were equipped with a backgate, which was also used to vary $n_e$ for measurements at 4.2 K.) For the present measurements, $V_0$ was nearly independent of $n_e$. Note that $V_0$ is only a few percent of the Fermi energy $E_\mathrm{F}$.

Figure \ref{SmoothBG} illustrates a typical example of the new low-field magnetoresistance oscillation. Solid curve in the main panel represents magnetoresistance $\Delta\rho_{xx}/\rho_0$ taken at 1.4 K by standard ac lock-in technique (100 nA, 70 Hz), showing PMR and CO\@. Close inspection reveals small amplitude oscillations, superposed on PMR background having overwhelmingly larger magnitude, at low fields ($|B|$$\alt$0.03 T). Subtraction of the slowly varying background component (the dotted curve in the main panel) results in the black solid curve in the inset, which clearly shows the oscillation. Alternatively and more conveniently, numerical differentiation with respect to the magnetic field may be taken. The red dashed trace in the inset shows the second derivative $d^2(\Delta\rho_{xx}/\rho_0)/dB^2$. By comparing solid and dashed traces, it is recognized that minima in $d^2(\Delta\rho_{xx}/\rho_0)/dB^2$ corresponds to maxima in $\Delta\rho_{xx}/\rho_0$, reminiscent of the case in sinusoidal oscillation.

\begin{figure}[bt]
\includegraphics[bbllx=40,bblly=280,bburx=800,bbury=800,width=8.5cm]{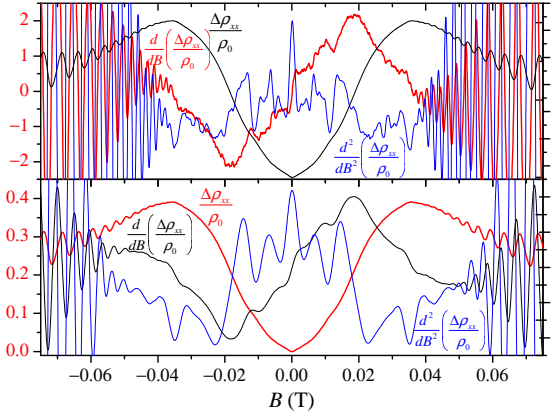}%
\caption{\label{meascmp}Top panel: $\frac{d}{dB}(\frac{\Delta\rho_{xx}}{\rho_0})$ directly measured by double lock-in technique. Numerically integrated $\frac{\Delta\rho_{xx}}{\rho_0}$ and differentiated $\frac{d^2}{dB^2}(\frac{\Delta\rho_{xx}}{\rho_0})$ are also shown. Bottom panel: magnetoresistance measured by standard lock-in technique with numerically differentiated $\frac{d}{dB}(\frac{\Delta\rho_{xx}}{\rho_0})$ and $\frac{d^2}{dB^2}(\frac{\Delta\rho_{xx}}{\rho_0})$. The red curves represent the raw data. These data were taken on sample \#2 at 1.4 K\@.}
\end{figure}

To confirm that the oscillatory features are by no means an artefact due to data processing (background subtraction or numerical differentiation), we measured $d(\Delta\rho_{xx}/\rho_0)/dB$ directly by employing double lock-in technique: small amplitude ac modulation $B_\mathrm{mod}$=1 mT (7 Hz) was superposed to the dc $B$ sweep during the standard lock-in (70 Hz) magnetoresistance measurement and the component of the lock-in-amplifier output that follows $B_\mathrm{mod}$ was recorded, which should then be proportional to $d(\Delta\rho_{xx}/\rho_0)/dB$. The first derivative thus acquired is compared in Fig.\ \ref{meascmp} with the numerically differentiated $d(\Delta\rho_{xx}/\rho_0)/dB$ obtained from the standard measurement. This, as well as other numerically integrated or differentiated traces shown for both types of measurements, establishes the consistency of the results obtained by the two different types of measurement. In the following, we use the data acquired by standard measurements and their numerical differentiation.

\begin{figure}[t]
\includegraphics[bbllx=30,bblly=280,bburx=660,bbury=800,width=8.5cm]{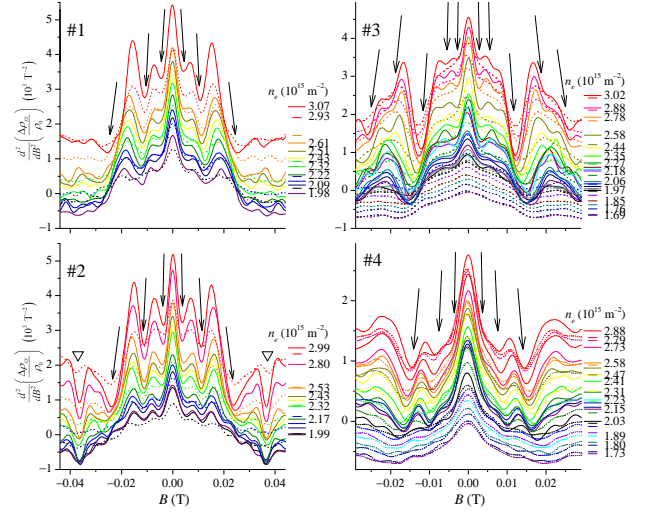}\\%
\caption{\label{d2dGs} Second derivative traces for various values of $n_e$ for samples \#1--\#4. Solid and dotted curves are for 1.4 K and 4.2 K, respectively. Traces are offset by an amount proportional to the change in $n_e$ due to illumination or backgate bias. Bases of the offset traces are noted by long (1.4 K) or short (4.2 K) horizontal lines on the right, with (selected) values of $n_e$ (in 10$^{15}$ m$^{-2}$). Arrows indicate the positions of minima picked out to be plotted in Fig.\ \ref{Bmins}. Triangles point to minima that do not shift with $n_e$ observed only for \#2.}
\end{figure}

\begin{figure}[tb]
\includegraphics[bbllx=30,bblly=280,bburx=850,bbury=850,width=8.5cm]{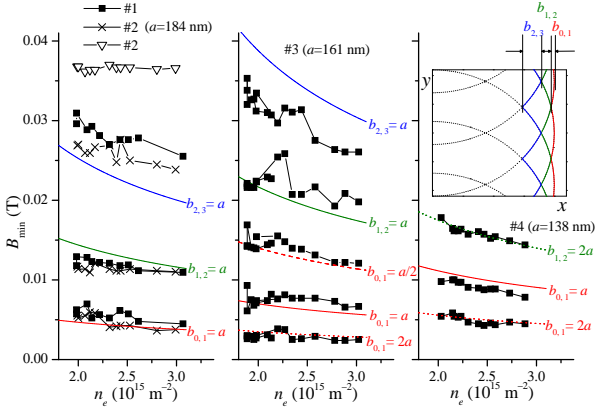}%
\caption{\label{Bmins}Minima positions of $\frac{d^2}{dB^2}(\frac{\Delta\rho_{xx}}{\rho_0})$ for samples \#1--\#4. Curves of positions expected from Eq.\ (\ref{calcBmin}) are also shown. Inset depicts the open orbits.}
\end{figure}

Figure \ref{d2dGs} shows $d^2(\Delta\rho_{xx}/\rho_0)/dB^2$ traces for samples \#1--\#4 for various values of $n_e$. Solid and dotted curves represent measurement at 1.4 K and 4.2 K, respectively. The followings can be read off from the figures: the oscillation becomes more prominent (i) with increasing $n_e$, and (ii) with decreasing temperature; the positions of minima (iii) shift to lower-field side with increasing  $n_e$, but (iv) do not depend on temperature, and (v) are not very sensitive to $V_0$ as long as $a$ remains unchanged (by comparing \#1 and \#2). Only sample \#2 displays extra minima at $|B|$$\sim$0.037 T (indicated by downward triangles) that do not shift with $n_e$. We defer discussing the last stationary minima for a moment. The features (i) and (ii) are likely to be a consequence of improved mobility by increasing $n_e$ or by lowering temperature. It is worth pointing out that (iii) is in marked contrast with the case for CO, where the ratio $R_c/a$ is preserved for extrema and therefore their positions shift to higher field side with $n_e$ in proportion to $k_\mathrm{F}$ ($\propto$$\sqrt{2\pi n_e}$). For more quantitative understanding, the positions of minima $B_\mathrm{min}$, marked by arrows and triangles in Fig.\ \ref{d2dGs}, are plotted in Fig.\ \ref{Bmins} (only $B_\mathrm{min}$$>$0 is shown). Replot of Fig.\ \ref{Bmins} with the ordinate $B_\mathrm{min}$ replaced with $B_\mathrm{min}\sqrt{n_e}$ (not shown) reveals that approximately $B_\mathrm{min}$$\propto$$n_e^{-1/2}$$\propto$$k_\mathrm{F}^{-1}$. In what follows, we describe our interpretation of the new oscillation that explains both the $n_e$- and $a$-dependence of $B_\mathrm{min}$.

\begin{figure}[tb]
\includegraphics[bbllx=50,bblly=50,bburx=580,bbury=280,width=8.5cm]{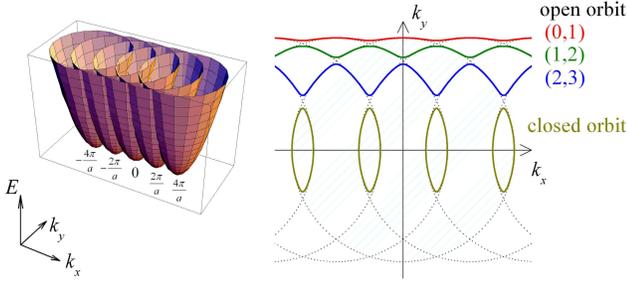}%
\caption{\label{Fermi}Left: A sketch of 2D band diagram of ULSL in the NFE approximation. A small gap opens up where two paraboloids intersect. Right: Illustration of the Fermi surface (cross section of the left figure by $E$=$E_\mathrm{F}$ plane). The shaded circle represents a Fermi circle were it not for the Bragg reflection. Open and closed orbits are shown. An open orbit is dubbed ($j$,$k$), when it is made up by Bragg reflections between $j$-th and $k$-th nearest Fermi circles.}
\end{figure}

Repeated-zone-scheme band diagram of ULSL in $k_x$-$k_y$-$E$ space, in the nearly-free-electron (NFE) approximation, is shown in the left panel of Fig.\ \ref{Fermi}. Paraboloids are periodically placed along the $k_x$-axis with an interval of $2\pi/a$, and a small gap opens up at the intersections. Non-parabolicity is neglected, which is justified in GaAs-based 2DEG\@. For transport properties at low temperatures, it is sufficient to consider only the vicinity of the Fermi energy $E_\mathrm{F}$. The Fermi contour is shown in the right panel. Bragg reflections from the periodic potential reconstruct the Fermi contour from Fermi circles (exemplified by the shaded circle) into sets of open and closed orbits. The open orbit situated between $j$-th and $k$-th intersection points between Fermi circles, the points where Bragg reflection takes place (numbers are assigned sequentially from outside, i.e., from larger $|k_y|$, and 0 denotes $|k_y|$=$k_\mathrm{F}$), is indicated as ($j,k$). In a weak enough magnetic field $B$ where magnetic breakdown is negligible, electron trajectory in the reciprocal space tracks one of these orbits in the direction determined by the sign of $B$. The corresponding trajectories in real space is obtained after a rotation by $\pi$/2 and multiplication by the factor $\hbar/eB$=$\ell^2$. Therefore the open orbit ($j,k$) corresponds to the electron trajectory that runs in $y$ direction (parallel to the grating, see the inset in Fig.\ \ref{Bmins}), with width (neglecting $\eta$=$V_0/E_\mathrm{F}$$\alt$0.05),
\begin{equation}
b_{j,k}=\frac{\hbar k_\mathrm{F}}{e|B|}\left[\sqrt{1-\left(\frac{j\pi}{ak_\mathrm{F}}\right)^2}-\sqrt{1-\left(\frac{k\pi}{ak_\mathrm{F}}\right)^2}\right].
\label{width}
\end{equation}
The orbits can intuitively be viewed as those that trace one or two segments of cyclotron orbit repeatedly, diffracted by crystal momentum of the superlattice. Similar to cyclotron orbits, the dimension (width) of the orbits are inversely proportional to $B$. Substituting the sample parameters into Eq.\ (\ref{width}) reveals that $b_{j,k}$ becomes close to $a$ in the the magnetic field range of our present interest. It is then natural to assume that the transport properties are altered at the magnetic field where $b_{j,k}$ coincides with the multiples of $a$. We further assume that the open orbit enhances $\sigma_{yy}$ at the resonant condition $b_{j,k}$=$na$. The increment $\delta\sigma_{yy}$ will increase $\rho_{xx}$ by $\delta\rho_{xx}/\rho_0$$\simeq$$(B\mu)^2\delta\sigma_{yy}/\sigma_0$. Thus the magnetic field $B_\mathrm{min}$ for maxima in $\rho_{xx}$ (minima in $d^2(\Delta\rho_{xx}/\rho_0)/dB^2$) is given by
\begin{equation}
\left|B_\mathrm{min}^{j,k,n}\right|=\frac{\hbar k_\mathrm{F}}{nae}\left[\sqrt{1-\left(\frac{j\pi}{ak_\mathrm{F}}\right)^2}-\sqrt{1-\left(\frac{k\pi}{ak_\mathrm{F}}\right)^2}\right].
\label{calcBmin}
\end{equation}
The $B_\mathrm{min}$'s calculated by Eq.\ (\ref{calcBmin}) are displayed in Fig.\ \ref{Bmins}, showing reasonable agreement with experimental minima \cite{half}. Note that for small enough $(\pi/ak_\mathrm{F})^2$, $B_\mathrm{min}^{j,k,n}$$\simeq$$(k^2-j^2)\pi h/(4 n e a^3 k_\mathrm{F})$, explaining the observed approximate $n_e$$^{-1/2}$ dependence.

Similar geometric resonance of open orbits was theoretically considered long time ago in the context of magnetoacoustic attenuation in metals \cite{Sievert67}. In that case, the width of open orbits becomes resonant with the wavelength of ultrasonic waves. In the present case, it is the superlattice that generates the open orbits in the first place, that also works as the ``resonator''. In principle, equivalent mechanism can also be operative in metals. However, it will require prohibitively large magnetic field ($\sim$100 T) because of approximately $B_\mathrm{min}$$\propto$$a^{-3}k_\mathrm{F}^{-1}$ dependence, since the lattice constant of metals are more than two orders of magnitude smaller than our $a$ (although $k_\mathrm{F}$ is roughly two orders of magnitude larger). The situation is analogous to the quest for Hofstadter butterfly \cite{Albrecht01}, where large artificial lattice constant reduces the required magnetic field into experimentally attainable range.

Now we turn to the minima in sample \#2 that do not shift with $n_e$ (open triangles in Figs.\ \ref{d2dGs} and \ref{Bmins}). As mentioned earlier, the grating of the sample \#2 is periodically ($a^\prime$=575 nm) notched along its length to relax strain. This inevitably introduces periodic modulation along the grating ($y$ direction), although it is designed to be negligible ($\sim$0.015 meV estimated by measuring ULSL with $a$=92 nm, twice the width of the notch \cite{Endo01c}) compared to $V_0$. Since open orbits have periodicity in the $y$ direction (see the inset to Fig.\ \ref{Bmins}), the period $(\hbar/eB)(2\pi/a)$ also can cause resonance when it equals $a^\prime$. The resonant field $B$=$(h/e)/aa^\prime$ is independent of $n_e$ and is $\sim$0.039 T for the sample, reasonably accounting for the observed minima. Analogous resonance is known again for magnetoacoustic attenuation \cite{Sievert67,Gavenda62,Tepley65} when supersonic wave propagate parallel to the open orbit.

So far we have assumed that $B$ is small enough so that magnetic breakdown do not exterminate the open orbits. This can be verified by a formula for breakdown probability \cite{Streda90} $p$=$\exp(-B_\mathrm{bd}/B)$ with
\begin{equation}
B_\mathrm{bd}=\frac{\pi m^* V_0^2}{8e\hbar E_\mathrm{F}}\left(\frac{\pi}{ak_\mathrm{F}}\right)^{-1}\left[1-\left(\frac{\pi}{ak_\mathrm{F}}\right)^2\right]^{-1/2}.
\end{equation}
The calculated largest $p$ for $B_\mathrm{min}$ in Fig.\ \ref{Bmins} are 0.80, 0.83, 0.93, and 0.95 for samples \#1, 2, 3, and 4 respectively. Therefore there still remains measurable probability $1-p$ for open orbits to survive breakdown, which supports our present interpretation.

In a recent paper \cite{Deutschmann01}, Deutschmann and coworkers reported $B^{-1}$ periodic magnetoresistance oscillation attributable to self-interference along closed orbits, equivalent to Shubnikov de Haas (SdH) oscillation for Fermi circles. In our samples, however, no $B^{-1}$ oscillation other than SdH was observed. The discrepancy can be traced back to the difference in the magnitude of $V_0$. In Ref.\ \cite{Deutschmann01}, the ULSL sample was fabricated by a sophisticated technique called the cleaved-edge overgrowth, which enables to meet two hardly compatible requirements of small $a$ (100 nm) and large $V_0$ ($\eta$$\agt$0.1). Much smaller $V_0$ in our sample makes breakdown probability close to unity in the magnetic field range where quantum oscillations are observable ($|B|$$\agt$0.1 T), letting the quantum oscillation be dominated by that from the orbits with full breakdown, namely, the Fermi circles, which is simply the SdH oscillation.

Finally, we discuss the prerequisites for observing the geometric resonance of open orbits. First of all, for the miniband effects not to be obscured, the number of minibands, $ak_\mathrm{F}/\pi$, below $E_\mathrm{F}$ should not be too large. The number in the present study ranges 4.5--8. Secondly, since miniband gap is small owing to small $V_0$, the gap can easily be collapsed by disorder, e.g., in the duty ratio of the grating. Therefore it seems necessary to minimize such disorder. We conjecture that the simple process we adopted for introducing potential modulation (no etchings or lift-offs) and the choice of electron-beam resist with the potentiality of sub-10 nm resolution \cite{Fujita96} were advantageous in this respect. Thirdly, as pointed out earlier, $B_\mathrm{min}$ rapidly decreases with $a$. This also sets an upper limit for $a$, since minima will be unresolvable because of the factor $\propto$$B^2$ in $\delta\rho_{xx}$ if $B_\mathrm{min}$ is too small. The same reason explains why $b_{0,1}$=2$a$ observed for samples \#3, 4 were not clearly resolved for samples \#1, 2. On the other hand, for open orbits to pull through the magnetic breakdown, $B_\mathrm{min}$ should not be too large, i.e., $a$ should not be too small unless $V_0$ can be simultaneously made large. In fact, we were unable to find the low-field oscillation in ULSL samples with $a$=115 nm and 92 nm. It appears that our choice of $a$ fortuitously placed us within a small window for observing the phenomena.

To summarize we have observed a new class of small amplitude magnetoresistance oscillation in ULSL, in the low-field regime where magnetic breakdown is still not prevailing. The oscillation is attributed to geometric resonance between the width of open orbits and the period $a$ of ULSL or, in the case the ULSL contains also a perturbing periodic modulation $a^\prime$ along the grating, between the period of open orbit and $a^\prime$.

This work was supported by Grant-in-Aid for Scientific Research (C) (15540305) and (A) (13304025) and Grant-in-Aid for COE Research (12CE2004) from Ministry of Education, Culture, Sports, Science and Technology.

\bibliography{lsls,ourpps,misc,noteNewO}

\begin{thebibliography}{16}
\bibitem[{hal()}]{half}
\bibinfo{note}{Sample \#3 includes $B_\mathrm{min}$ assigned as
  $b_{0,1}$=$a$/2, which has no equivalent in other samples, and has much
  larger width and amplitude than other minima for the same sample (see Fig.\
  \ref{d2dGs}). It seems rather uncertain whether this particular
  $B_\mathrm{min}$ may be interpreted on the same footing as others}.
\end{thebibliography}

\end{document}